\documentclass[aps,10pt,tighten,twocolumn]{revtex4}%
\usepackage{amssymb}
\usepackage{graphicx}
\usepackage{amsmath}
\usepackage{amsfonts}%
\providecommand{\U}[1]{\protect\rule{.1in}{.1in}}
\providecommand{\U}[1]{\protect\rule{.1in}{.1in}}
\providecommand{\U}[1]{\protect\rule{.1in}{.1in}}
\providecommand{\U}[1]{\protect\rule{.1in}{.1in}}
\providecommand{\U}[1]{\protect\rule{.1in}{.1in}}

\begin{document}
\title[ ]{Vertex Corrections on the Anomalous Hall Effect in Spin-polarized
Two-dimensional Electron Gases with Rashba Spin-orbit Interaction}
\author{Jun-ichiro Inoue}
\email{inoue@nuap.nagoya-u.ac.jp}
\affiliation{Department of Applied Physics, Nagoya University, Nagoya 464-8603, Japan}
\author{Takashi Kato, Yasuhito Ishikawa, and Hiroyoshi Itoh}
\affiliation{Department of Applied Physics, Nagoya University, Nagoya 464-8603, Japan}
\author{Gerrit E. W. Bauer}
\affiliation{Kavli Institute of NanoScience, Delft University of Technology, Lorentzweg 1,
2628CJ Delft, The Netherlands}
\author{Laurens W. Molenkamp}
\affiliation{Physikalisches Institut (EP3), Universit\"{a}t W\"{u}rzburg, D-97074
W\"{u}rzburg, Germany}
\pacs{PACS number: }

\begin{abstract}
We study the effect of disorder on the intrinsic anomalous Hall (AH)
conductivity in a spin-polarized two-dimensional electron gas with a
Rashba-type spin-orbit interaction. We find that AH conductivity vanishes
unless the lifetime is spin-dependent, similar to the spin Hall (SH)
conductivity in the non-magnetic system. In addition, we find that the SH
conductivity does not vanish in the presence of magnetic scatterers. We show
that the SH conductivity can be controlled by changing the amount of the
magnetic impurities.

\end{abstract}
\maketitle

The spin-orbit interaction (SOI) in semiconductors allows optical and
electrical control of spins, and because of this has recently attracted much
attention in the field of spintronics. The SOI gives rise to unusual Hall
effects, such as the anomalous Hall (AH) effect \cite{hall2} in ferromagnets,
and the spin Hall (SH) effect in normal conductors
\cite{dyakonov,hirsh,murakami,sinova}. The detailed mechanisms of these
effects are still controversially discussed. For the AH effect, originally an
intrinsic (i.e., band-structure induced) mechanism originating from an
effective magnetic field in momentum space was put forward \cite{karplus},
followed by extrinsic mechanisms, referred to as skew \cite{smit} and
side-jump \cite{berger} scattering at impurities. Most experiments have been
interpreted in terms of the extrinsic mechanisms, but the intrinsic AH effect
has recently been shown to quantitatively explain the AH effect in
ferromagnetic semiconductors \cite{jungwirth,ohno,yao}. While a current bias
does not excite a Hall voltage in normal metals at zero external field, a Hall
effect of spin currents should persist in the presence of intrinsic or
extrinsic SOI. This spin Hall effect was originally predicted assuming
extrinsic scattering \cite{dyakonov,hirsh}, but, analogous to the AH effect,
an intrinsic SH effect is also possible. Murakami \textit{et al}%
.\cite{murakami} predicted that the effective magnetic field associated to the
Berry phase in the valence band induce drift of up and down spin carriers
towards opposite directions in $p$-doped zincblende-type semiconductors.
Sinova \textit{et al}.\cite{sinova} predicted a universal spin Hall
conductivity for the two-dimensional electron gas (2DEG) with a Rashba-type
SOI produced by the asymmetry of the potential.

Recently, two groups \cite{kato,wunderlich} reported optical detection of spin
accumulation of opposite signs at the sample edges in current-biased
nonmagnetic semiconductors. Such behavior can be caused by the extrinsic SH
effect, as shown explicitly by Ref. \cite{SZhang} and is believed to be a
signature of the intrinsic SH effect as well. However,
the interpretation of the experimental results is not straightforward.

Current theories predict that the intrinsic SH effect can strongly be
suppressed by disorder effects. Especially, the intrinsic SH current vanishes
identically by disorder scattering of electrons in the bulk of a Rashba-split
2DEG \cite{inoue1,raimondi,rashba}. On the other hand, the Rashba-split 2DEG
is believed to be rather special in this respect, since in other systems the
intrinsic SH effects survives disorder scattering. This has been demonstrated
for 2DEGs with an SO\ interaction that is not directly proportional to the
wave vector, and for three dimensional systems in the presence of the
Dresselhaus SO interaction, or a Luttinger type of SO interaction in the
valence band\cite{murakami2,malshukov,bernevig2}.

Since one may view the SH effect as the zero-magnetization limit of the AH
effect, the vertex-canceling of the intrinsic SH in a Rashba-split 2DEG raises
the issue whether or not the intrinsic AH effect is affected equally strongly
by disorder scattering. And can, \textit{vice versa}, the intrinsic SH effect
survive in a disordered system when magnetic effects are added, \textit{e.g}.,
carrier scattering by magnetic impurities? \ These are the questions we
address in this Letter. We focus on a disordered 2DEG with the intrinsic
Rashba-type SOI and investigate the effects of a constant exchange potential
as well magnetic impurities on the AH and SH effects by generalizing the
method we applied previously \cite{inoue1,inoue2} to calculate the
conductivity in the diffusive transport regime.

We will show that the intrinsic AH conductivity in Rashba split 2DEG with
uniform spin polarization vanishes unless the life-time is spin-dependent,
which is correct up to the second order of the SOI. This exemplifies the
strong similarity between SH and AH effects. We also find that the SH
conductivity is non-zero in the presence of magnetic impurity scattering.
These results are not only relevant for a full understanding of the Hall
effects \cite{bauer,inoue3}, but should also apply to high g-factor, high
mobility narrow gap magnetic semiconductors such as (Hg,Mn)Te that exhibit
many of the features needed for a clean observation of the intrinsic SOI
induced Hall effects\cite{HgTe} in a transport experiment.

We start with the Hamiltonian
\begin{equation}
H=H_{0}+V_{m},
\end{equation}
where $H_{0}$ is the unperturbed Hamiltonian for the 2DEG and $V_{m}$ the
random potential caused by impurities. For the non-magnetic 2DEG, $H_{0}$
equals the Rashba Hamiltonian $H_{R}$. In Pauli-spin and momentum space we
have
\begin{equation}
H_{R}=\left(
\begin{array}
[c]{cc}%
\frac{\hbar^{2}}{2m}k^{2} & i\lambda\hbar k_{-}\\
-i\lambda\hbar k_{+} & \frac{\hbar^{2}}{2m}k^{2}%
\end{array}
\right)  ,
\end{equation}
where $k_{\pm}=k_{x}\pm ik_{y}\ $with in-plane momentum vector $\mathbf{k=}%
\left(  k_{x},k_{y}\right)  $ and $\lambda$ is the tunable strength of the
spin-orbit coupling. The eigenvalues and eigenfunctions of $H_{R}$\ are
\begin{equation}
E_{k\pm}=\frac{\hbar^{2}k^{2}}{2m}\pm\lambda\hbar k,
\end{equation}
and%
\begin{equation}
|s=\pm\rangle=\left(
\begin{array}
[c]{c}%
isk_{-}/k\\
1
\end{array}
\right)  ,
\end{equation}
respectively.

In order to deal with the AH and SH conductivities in the diffusive transport
regime, we modify the Hamiltonian accordingly. In the calculation of the AH
conductivity, the unperturbed part of the Hamiltonian is given as%
\begin{equation}
H_{0}=H_{R}+\Delta_{ex}\sigma_{z},
\end{equation}
where the second part expresses an exchange potential, where $\sigma_{z}$ is
the $z$-component of the Pauli spin matrix. The random potential is supposed
to be short-ranged and isotropic, but may be spin-dependent:%
\begin{equation}
V_{m}=\sum_{i\sigma=\uparrow,\downarrow}V_{\sigma}\delta(\mathbf{r}%
-\mathbf{R}_{i}). \label{VAHE}%
\end{equation}
In the calculation of the SH conductivity, on the other hand, the unperturbed
Hamiltonian is $H_{R}$, but we incorporate magnetic impurities that give rise
to spin-flip scattering of electrons:
\begin{equation}
V_{m}=2J\sum_{i}\mathbf{s}\cdot\mathbf{S}\delta(\mathbf{r}-\mathbf{R}_{i}),
\label{VSHE}%
\end{equation}
where $\mathbf{s}$ and $\mathbf{S}$ are the spin operators of the conduction
electrons and the localized magnetic impurities, respectively, and $J$ is
their exchange coupling. We consider the regime in which the Kondo effect is
not important.

We compute the transport properties via the Kubo formula, using the simplest
possible definition of the charge current operator
\begin{equation}
J_{x(y)}=e\left[  (\hbar/m)k_{x(y)}\mathbf{1}-(+)\lambda\sigma_{y(x)}\right]
.
\end{equation}
In the diffusive regime it is convenient to include the SOI in the eigenstates
of the Hamiltonian, and treat the impurity potentials as perturbation. We can
then proceed to obtain the AH and SH conductivities as before by adopting the
Born approximation for the self-energy and the ladder approximation for the
current vertex that is determined self-consistently in order to satisfy the
Ward identity.

After a lengthy but straightforward calculation along the lines of Refs.
\cite{inoue1} and \cite{inoue2}, we have obtained the full expression (not
shown) for the AH conductivity in terms of retarded and advanced Green
functions in the Pauli spin space. The momentum integration of the products of
retarded and advanced Green function has been done analytically by assuming
$\lambda\ll\Delta_{ex}$ and keeping terms up to $\lambda^{2}$ and in the clean
limit with life-time broadening being smaller than the Rashba-splitting of the bands.

By ignoring the real part of the self-energy, $\Sigma_{\sigma}=\left(
n/L^{2}\right)  \sum_{k}|V_{\sigma}^{2}|g_{k\sigma}$, where $n/L^{2}$ is the
impurity density, and with lifetime $\tau_{\sigma}=\hbar/2|\Sigma_{\sigma}|$
we obtain
\begin{align}
\sigma_{yx}^{AH}  &  =\frac{4e^{2}\lambda^{2}D_{0}}{4\Delta_{ex}^{2}+\left(
|\Sigma_{\uparrow}|+|\Sigma_{\downarrow}|\right)  ^{2}}\times\left[
-\epsilon_{F\uparrow}|\Sigma_{\downarrow}|\tau_{\uparrow}+\epsilon
_{F\downarrow}|\Sigma_{\uparrow}|\tau_{\downarrow}\right. \nonumber\\
&  \left.  +\frac{\left(  \epsilon_{F\uparrow}\tau_{\uparrow}-\epsilon
_{F\downarrow}\tau_{\downarrow}\right)  ^{2}}{\left(  \tau_{\uparrow}%
\tau_{\downarrow}\right)  ^{1/2}}\frac{\Delta_{ex}\left(  |\Sigma_{\uparrow
}|+|\Sigma_{\downarrow}|+C\right)  }{4\Delta_{ex}^{2}+C^{2}}\right]  ,
\label{AHC}%
\end{align}
where $D_{0}=m_{e}/2\pi\hbar^{2}$ is the 2DEG density of states and%
\begin{align}
\epsilon_{F\uparrow(\downarrow)}  &  =\epsilon_{F}+(-)\Delta_{ex},\\
C  &  =\pi nD_{0}\left(  V_{\uparrow}-V_{\downarrow}\right)  ^{2}=\left(
\sqrt{|\Sigma_{\uparrow}|}-\sqrt{|\Sigma_{\downarrow}|}\right)  ^{2}.
\end{align}
The first two terms inside the square brackets in Eq. (\ref{AHC}) are the
non-vertex (bubble) part, and the last originates from the vertex correction.

We can simplify the expression for $\sigma_{yx}^{AH}$ further, assuming that
$\Delta_{ex}\gg|\Sigma_{\uparrow}|,|\Sigma_{\downarrow}|,$ $\tau
_{\uparrow(\downarrow)}=\tau\left(  1+(-)\delta\right)  $ with $\delta\ll1$,
neglecting the self-energy part in the denominator, and expanding up to
$\delta^{2}$ we have
\begin{equation}
\sigma_{yx}^{AH}\sim\sigma_{xx}^{SO}\left(  \frac{\epsilon_{F}}{\Delta_{ex}%
}\right)  ^{2}\frac{|\Sigma|}{\Delta_{ex}}\delta^{2},
\end{equation}
where $\sigma_{xx}^{SO}=2e^{2}\tau D_{0}\lambda^{2}$ is the correction to the
longitudinal conductivity by the SOI in a presence of non-magnetic impurities,
and $\tau=\hbar/2|\Sigma|$.

This is our main result for the AH conductivity. We can draw several
conclusions from the above expression. First, we very generally find that
$\sigma_{yx}^{AH}=0$ when the carrier lifetime is spin-independent. This also
follows from inspection of Eq. (\ref{AHC}) and we have verified that this
statement is valid even when the real part of the self-energy is retained. The
cancellation of the bubble by the vertex part in this limit is realized only
after self-consistent renormalization of the latter. We note the similarity
with the SH conductivity which vanishes by introducing electron scattering by
nonmagnetic impurities. One should note, however, again in analogy with the SH
effect, that the AH effect does not have to vanish for ferromagnets with other
band structures.

Secondly, the AH conductivity is proportional to $\lambda^{2}$ and independent
of the life time $\tau$ (note that $\tau|\Sigma|$ in eq. (12) is independent
of the life-time), very similar to the extrinsic AH conductivity based on the
side-jump mechanism \cite{nagaosa}. Our expression is not identical that for
side-jump scattering, however; for example, Crepieux and Bruno \cite{crepieux}
obtained a non-zero $\sigma_{yx}^{AH,sidejump}$ even in the limit
$\tau_{\uparrow}=\tau_{\downarrow}$ with a constant magnetic moment $M$. Skew
scattering does not occur in our model since the SOI is intrinsic and
homogenous, and carrier scattering is caused by an impurity potential that
does not contain SOI.

Finally, the AH conductivity is odd with respect to the magnetization $M$,
since $\Delta_{ex}\varpropto M$. Therefore, the AH conductivity changes sign
when the magnetization is reversed, as expected. We see in Eq. (12)
$\sigma_{yx}^{AH}$ increases with decreasing $M$. Although Eq. (12) may not be
strictly valid for $M\rightarrow0$ because we assumed $\Delta\gg
|\Sigma_{\uparrow}|,|\Sigma_{\downarrow}|$, $\sigma_{yx}^{AH}$ has to vanish
in this limit. Thus we expect a non-monotonic dependence of $\sigma_{yx}$ on
$M$.

We now turn to magnetic impurity effects on the SH conductivity. To this end,
we will introduce spin-dependent random potentials into Rashba-split 2DEG. The
random potentials of $V_{\sigma}$ are not suitable for the present purpose,
since $\langle V_{\sigma}\rangle$ is spin-dependent which gives rise to a
finite exchange potential (and quite trivially a finite SH effect) in the
lowest order approximation. We therefore adopt an $s-d$ type interaction
$\delta(\mathbf{r}-\mathbf{R}_{i})\mathbf{s}_{i}\cdot\mathbf{S}_{i}$\ between
conduction electron $\mathbf{s}_{i}$\ and localized impurity $\mathbf{S}_{i}$
spins\ on a magnetic impurity at $\mathbf{R}_{i}$. This type of interaction is
isotropic, and no first order correction appears. This isotropy persists in
the presence of the Rashba SOI, so the self-energy and Green function are
diagonal in\ the $|s=\pm\rangle$ space which is a convenient basis. The
self-energy in the Born approximation is proportional to the thermal average
of a localized spin $\langle S^{2}\rangle$, which consists of three
contributions, one from spin-conserving scattering, $\langle\sigma_{z}%
S_{z}\sigma_{z}S_{z}\rangle,$ and two from spin-flip scattering,
$\langle\sigma_{+}S_{-}\sigma_{-}S_{+}\rangle$ and $\langle\sigma_{-}%
S_{+}\sigma_{+}S_{-}\rangle$. Accordingly, the self-energy
\begin{equation}
|\Sigma|\simeq\frac{\hbar}{2}\left(  \frac{1}{\tau_{0}}+\frac{1}{\tau_{sf}%
}\right)  ,
\end{equation}
is governed by two scattering rates. $1/\tau_{0}$ and $1/\tau_{sf}$ correspond
to the $\langle S_{z}^{2}\rangle$ and $\left(  1/2\right)  \langle S_{+}%
S_{-}+S_{-}S_{+}\rangle$ contributions, respectively. In our case,
$1/\tau_{sf}=2/\tau_{0}$ since the potentials are isotropic.

After evaluating the vertex corrections self-consistently, we may define an
effective current vertex as
\begin{equation}
\mathbf{\tilde{J}}_{x}^{kk}=e\left[  \frac{\hbar}{m}k_{x}\mathbf{1}%
+\frac{\lambda+\lambda^{\prime}}{k}\left(  k_{x}\mathbf{\sigma}_{z}%
-k_{y}\mathbf{\sigma}_{y}\right)  \right]  , \label{CV}%
\end{equation}
where $\lambda^{\prime}$ is an effective SOI originated from the vertex
corrections which is the solution of%
\begin{equation}
\lambda^{\prime}=-\langle\upsilon_{m}^{2}\rangle\left[  A+(\lambda
+\lambda^{\prime})(B+D)\right]  \label{ESOI}%
\end{equation}
with
\begin{equation}
\langle\upsilon_{m}^{2}\rangle=\frac{n_{m}J^{2}\langle S_{z}^{2}\rangle
}{4L^{2}}.
\end{equation}
where $n_{m}/L^{2}$ is the density of magnetic impurities. The expression of
the current vertex $\lambda^{\prime}$\ is apparently\cite{inoue1,inoue2} the
same for magnetic and nonmagnetic impurities except for the definition of
$\langle\upsilon_{m}^{2}\rangle$. $\ A,B,$ and $D$ are momentum integrals over
products over retarded and advanced Green functions that can be integrated
analytically, such that%
\begin{equation}
\lambda^{\prime}=\lambda\frac{\Delta_{0}}{8+7\Delta_{0}},
\end{equation}
with $\Delta_{0}=\left(  2\tau_{0}\lambda k_{F}\right)  ^{2}$ for magnetic
impurities. Here $\langle\upsilon_{m}^{2}\rangle$ has been inverted
to\ $\tau_{0}$ using the relation $\langle S_{z}^{2}\rangle=\langle
S^{2}\rangle/3$. Note that $\lambda^{\prime}=-\lambda$ for nonmagnetic
impurities\cite{inoue1}. The final result of the SH conductivity is the same
for magnetic and nonmagnetic impurities%
\begin{equation}
\sigma_{yx}^{\sigma_{z}}=\frac{e\left(  \lambda+\lambda^{\prime}\right)
}{4\pi\hbar\lambda}.
\end{equation}
We therefore find that the SH conductivity, which has been found before to
vanish for nonmagnetic impurities, becomes finite in the presence of magnetic
impurities. This is our main result concerning the SH effect.

There are two main reasons for the different SH conductivities for magnetic
and nonmagnetic impurities. One is the difference in sign in the matrix
elements of the random potentials. The matrix elements for the magnetic
potential include terms $\langle s=\pm|\mathbf{\sigma}_{z}|s=\pm\rangle$,
while in the nonmagnetic case the corresponding matrix element is $\langle
s=\pm|\mathbf{1}|s=\pm\rangle$ which turns out to have the opposite sign. The
other reason is the expression for $\langle\upsilon_{m}^{2}\rangle$, which is
proportional to $\langle S_{z}^{2}\rangle$, not $\langle S^{2}\rangle$ as in
the self-energy because the $\langle S_{+}S_{-}+S_{-}S_{+}\rangle$ term
couples with the linear terms of $k_{x}$ or $k_{y}$ and vanishes in the
momentum integral. Therefore, the vertex function includes only $\tau_{0}$,
while the self-energy basically includes both $\tau_{0}$ and $\tau_{sf}$.

The present results modify the expressions we obtained
previously\cite{inoue1,inoue2} for the longitudinal charge conductivity and
the in-plane spin, as follows
\begin{align}
\sigma_{xx}  &  =2e^{2}\tau_{0}\left[  \frac{n_{0}}{m}+D_{0}\lambda
^{2}-\lambda\left(  \lambda+\lambda^{\prime}\right)  \frac{D_{0}}{2}%
\frac{\Delta_{0}}{1+\Delta_{0}}\right]  ,\\
\langle s_{y}\rangle &  =4\pi eD_{0}\tau_{0}\left[  \lambda-(\lambda
+\lambda^{\prime})\frac{2+\Delta_{0}}{1+\Delta_{0}}\right]  E,
\end{align}
where $n_{0}$ is the carrier density, and $E$ is the applied electric field.
Since these observables do not vanish under the vertex correction even for
nonmagnetic impurities, the effect of modeling magnetic impurities is small.

Finally, let us consider the case where magnetic and non-magnetic impurities
coexist. Since there is no mixing between magnetic and nonmagnetic scattering,
the current vertex is still defined by Eq. (\ref{CV}), but now we have
\begin{equation}
\lambda^{\prime}=\lambda_{m}^{\prime}-\lambda_{n}^{\prime},
\end{equation}
where $\lambda_{m}^{\prime}$ and $\lambda_{n}^{\prime}$ are effective SOIs
caused by the magnetic and nonmagnetic impurities, respectively. Both satisfy
Eq. (\ref{ESOI}) with $\langle\upsilon_{m}^{2}\rangle$ for $\lambda
_{m}^{\prime}$ and with $\langle\upsilon_{n}^{2}\rangle=n_{n}\langle
V^{2}\rangle/4L^{2}$ for $\lambda_{n}^{\prime}$. Here, $n_{n}/L^{2}$ and $V$
are the density of nonmagnetic impurities and impurity potential,
respectively. Then $\lambda^{\prime}$ also satisfies the relation (\ref{ESOI})
with $\delta\langle\upsilon^{2}\rangle=\langle\upsilon_{m}^{2}\rangle-$
$\langle\upsilon_{n}^{2}\rangle$, which determines $\lambda^{\prime}$ self-consistently.

The self-energy is on the other hand given by the sum of the contributions
from magnetic and non-magnetic impurities, $|\Sigma|=|\Sigma_{m}|+|\Sigma
_{n}|\equiv\hbar/2\tau,$with
\begin{equation}
\frac{1}{\tau}=\left[  n_{m}4J^{2}\langle S^{2}\rangle+n_{n}\langle
V^{2}\rangle\right]  \frac{m}{\hbar^{3}}.
\end{equation}
Consider varying $n_{m}$ and $n_{n}$ such that the life time $\tau$ is kept
constant. Noting that $\langle S_{z}^{2}\rangle=\langle S^{2}\rangle/3$ in the
paramagnetic state, the effective spin-orbit interaction is then given as%
\begin{align}
\lambda^{\prime}  &  =\left[  n_{m}J^{2}\frac{4}{3}\langle S^{2}\rangle
-\frac{\hbar^{3}}{\tau m}\right]  \frac{m\tau}{2\hbar^{3}}\lambda\nonumber\\
&  \times\frac{\Delta_{0}}{1+\Delta_{0}+\left[  n_{m}J^{2}\frac{4}{3}\langle
S^{2}\rangle-\frac{\hbar^{3}}{\tau m}\right]  \frac{m\tau}{2\hbar^{3}}\left(
2+\Delta_{0}\right)  }.\nonumber\\
&
\end{align}
This expression reproduces the results $\lambda^{\prime}=-\lambda$ and
$\lambda^{\prime}=\lambda\Delta_{0}/(8+7\Delta_{0})$ for $n_{m}=0$ and
$n_{n}=0$, respectively. At an intermediate value of $n_{m}$, $\delta
\langle\upsilon^{2}\rangle$ can be zero. In this case, the vertex correction
itself vanishes, and the universal spin Hall conductivity is realized even in
the presence of impurity scattering.

These results shed some doubt on Rashba's general argument for a generally
vanishing SH conductivity \cite{rashba}. since he introduced a vector
potential by external magnetic field but neglected the Zeeman term. Indeed,
the latter gives rise to a non-vanishing spin Hall current even in Rashba
split 2DEG \cite{adagideli}. It should also be noted that the present results
have been obtained for an infinite system. Finite size systems may show a
non-vanishing SH conductance \cite{nikolic}\ or edge SH
currents\cite{adagideli,mishchenko}.

In conclusion, we studied the intrinsic AH effect in the diffusive transport
regime for a spin-polarized 2DEG with a Rashba-type SOI. We found that the AH
conductivity vanishes unless the life-time is spin-dependent, indicating a
strong similarity between the intrinsic AH and SH effects. Inclusion of a
spin-dependent lifetime yields a non-vanishing AH conductivity, the expression
of which is similar to that obtained by the side-jump mechanism. The SH
conductivity for the Rashba 2DEG with magnetic scattering of electrons has
found finite even in the diffusive transport regime. The SH conductivity in
the presence of both magnetic and nonmagnetic impurities shows that the SH
conductivity may be controlled by changing the amount of the magnetic impurities.

The authors acknowledge fruitful discussions with N. Nagaosa, H. Ohno, J.
Sinova, S. Murakami, and B. Nikolic. This work was supported by Grants-in-Aid
for Scientific Research in Priority Areas \textquotedblleft Semiconductor
Nanospintronics\textquotedblright\ of The Ministry of Education, Culture,
Sports, Science, and Technology of Japan, NAREGI Nanoscience Project,
Grant-in-Aid for the 21st Century COE "Frontiers of Computational Science",
the FOM Foundation, and the DFG (SFB 410).

\end{document}